\documentclass[11pt]{extarticle}
\usepackage[utf8]{inputenc}
\usepackage{amssymb,amsmath}
\usepackage[english]{babel}

\begin{document}

\textbf{Vaidya and generalized Vaidya solutions by gravitational decoupling} \\

\begin{center}
Vitalii Vertogradov $^{1}$, Maxim Misyura $^{2}$ \\

$^{1}$ \quad Physics department, Herzen state Pedagogical University of Russia, 48 Moika Emb., Saint Petersburg 191186, Russia \\
  SPB branch of SAO RAS; vdvertogradov@gmail.com\\
$^{2}$ \quad  Department of High Energy and Elementary Particles Physics,
  Saint Petersburg State University, University Embankment 7/9, Saint Petersburg, 199034, Russia\\
  Physics department, Herzen state Pedagogical University of Russia, 48 Moika Emb., Saint Petersburg 191186, Russia; 
max.misyura94@gmail.com
\end{center}

\abstract{In this paper, we apply the gravitational decoupling
method for dynamical systems in order to obtain a new type of solution
that can describe a hairy dynamical black hole. We consider three cases
of decoupling. The first one is the simplest and most well known when the
mass function is the function only of space coordinate $r$. The second
case is a Vaidya spacetime case when the mass function depends on time $v$
. Finally, the third case represents the generalization of these two cases:
the mass function is the function of both $r$ and $v$. We also calculate
the apparent horizon and singularity locations for all three cases.}


\section{Introduction}

Black holes are one of the most fascinating objects in our Universe.
Currently, we can make direct observations of them via detection of their
gravitational waves~\cite{bib:1, bib:2} or black hole
shadow~\cite{bib:3, bib:4}.

The famous no-hair theorem states that a black hole might have only three
charges: the mass $M$, angular momentum  $J$,  and~electric charges
$Q$~\cite{bib:5}.    However, it can be shown that black holes can have
other charges and there is so-called soft hair~\cite{bib:6}. Among~other
possibilities for evading the no-hair theorem is to use the gravitational
decoupling method~\cite{bib:gd1, bib:gd2, bib:gd3}.

It is well known that obtaining the analytical solution of the Einstein
equations is a difficult task in most cases. We know that we can
obtain an analytical solution of the spherically symmetric spacetime in
the case of the perfect fluid as the gravitational source. However, if~
we consider the more realistic case when the perfect fluid is coupled to
another matter, it is nearly impossible to obtain the analytical
solution. In~papers~\cite{bib:gd1, bib:gd2, bib:gd3}, it was shown using
the Minimal Geometric Deformation (MGD)~\cite{bib:7, bib:8} method
that we can decouple the gravitational sources, for~example, one can
write the energy-momentum tensor $T_{ik}$ as:
\begin{equation} \label{eq:thefirst}
T_{ik}=\tilde{T}_{ik}+\alpha\Theta_{ik} \,.
\end{equation}
where $\tilde{T}_{ik}$ is the energy-momentum tensor of the perfect
fluid and $\alpha$ is the coupling constant to the energy-momentum tensor
$\Theta_{ik}$. 
It is possible to solve Einstein's field equations for a gravitational
source whose energy-momentum tensor is expressed as
\eqref{eq:thefirst} by solving Einstein's field equations for each
component $\tilde{T}_{ik}$ and $\Theta_{ik}$ separately. Then, by~a
straightforward superposition of the two solutions, we obtain the
complete solution corresponding to the source $T_{ik}$. Since Einstein's
field equations are non-linear, the~MGD decoupling represents a novel
and useful method in the search for and analysis of solutions, especially
when we face scenarios beyond trivial cases, such as the interior of
stellar systems with gravitational sources more realistic than the ideal
perfect fluid, or~even when we consider alternative theories, which
usually introduce new features that are difficult to deal~with.

Moreover, there is only the gravitational interaction between
two sources, i.e.,
\begin{equation} \label{eq:conserv}
T^{ik}_{;k}=0 \rightarrow \tilde{T}^{ik}_{;k}=\alpha\Theta^{ik}_{;k}=0 \,.
\end{equation}

This fact allows us to think about $\Theta_{ik}$ as dark matter.
By applying the gravitational decoupling method, one can obtain
well-known black hole solutions with hair~\cite{bib:bh1, bib:bh2}.
However, this method is applied only to static or stationary cases. That
is, we obtain only an eternal hairy black hole solution. If~one
wants to understand the process of these hairy black hole formations,
then one should consider the gravitational collapse of the matter cloud. The~problem is that in the general case, the~(MGD) method is not
applicable due to its violation of condition \eqref{eq:conserv}
. The~
gravitational decoupling of a dynamical system is still a problem. One
of the first successful decouplings of the dynamical system was performed
in Ref.~\cite{bib:sharif}.

In this paper, we offer a model of the gravitational decoupling of
dynamical systems, which can be used to investigate the question of
gravitational collapse to a hairy black hole. By~using the hairy
Schwarzschild black hole solution obtained in Ref.~\cite{bib:bh1}, we introduce
the Eddington--Finkelstein coordinates in order to consider the non-zero
right hand side of the Einstein equations. In~this case, the~mass $M$ is
not a constant; however, it is the mass function of time $v$ and the radial
coordinate $r$
. As~a result, we obtained the Vaidya and generalized
Vaidya solutions. In~the Vaidya case, the~energy-momentum tensor
$\tilde{T}_{ik}$ represents the null dust. In~the generalized Vaidya case,
the $\tilde{T}_{ik}$ represents the mixture of two matter fields---type I
and type II~\cite{bib:hok, bib:vunk}

The Vaidya spacetime is the so-called radiating Schwarzschild
solution~\cite{bib:vaidya} and is one of the first examples of a
cosmic censorship conjecture violation~\cite{bib:pap}. The~Vaidya
spacetime is widely used in many astrophysical applications with strong
gravitational fields. In~general relativity, this spacetime assumed
added importance with the completion of the junction conditions at the
surface of the star by Santos~\cite{bib:santos1985non}. The~pressure at
the surface is non-zero, and~the star dissipates energy in the form of
heat flux. This made it possible to study dissipation and physical
features associated with gravitational collapse, as~shown by Herrera et
al.~\cite{bib:herrera2006some, bib:herrera2012dynamical,bib:joshivaidya}
. Some recent studies of the temperature properties inside the radiating
star include Reddy~et~al.~\cite{bib:reddy2015impact}, Thirukkanesh~et~al.~\cite{bib:thirukkanesh2012final}, and~Thirukkanesh and Govender~\cite{bib:thirukkanesh2013}. The~metric in Ref.~\cite{bib:Lindquist_1965} may
be extended to include both null dust and null string fluids leading to
the generalized Vaidya spacetime. The~properties of the generalized
Vaidya metric have been studied by Hussain~\cite{bib:husain1996exact},
Wang and Wu~\cite{bib:vunk}, and~Glass and Krisch~\cite{bib:Radiationstring1998, bib:twofluidatm1999}. Maharaj~et~al.~\cite{bib:Radiatingstars2012, bib:r5} modeled 
 a radiating star with a
generalized Vaidya atmosphere in general relativity. A~detailed study of
continual gravitational collapse of these spacetimes in the context of
the cosmic censorship conjecture was performed in Refs.~\cite{bib:Maombi1,
 bib:Maombi2, bib:ver1, bib:ver2, bib:ver3, bib:r2}. In~the geometrical context,
gravitational collapse has been considered in Lovelock gravity theory~\cite{bib:Gravitationalcollapse2011}, black holes in dynamical cosmology
backgrounds~\cite{bib:SurroundedVaidya2018}, and~in electromagnetic
fluids~\cite{bib:Heydarzvaidya2018}. The~influences of dust, radiation,
quintessence, and~the cosmological constant are included in these studies
. The~conformal symmetries and embedding properties of the generalized
Vaidya metric were studied in Refs.~\cite{bib:maharaj, bib:r4}. Other
properties of this spacetimes can be found in Refs.~\cite{bib:r1, bib:r3}.

The paper is organized as follows: In Section~\ref{sec2} we introduce the
Eddington--Finkelstein coordinates for the hairy Schwarzschild metric and
consider the solution of the Einstein equation with mass as the function
of the radial coordinate $r$ only. In~Section~\ref{sec3}, we obtain the hairy Vaidya
solution by solving the Einstein equations with mass depending on the
time $v$ only and calculate the apparent horizon for this metric. In~Section~\ref{sec4}, the~general case $M=M(v,r)$ is considered in order to obtain the
generalized hairy Vaidya spacetime, and~the apparent horizon, in~this
case, is also calculated. Section~\ref{sec5} is the~conclusion.

\section{The Perfect Fluid Case 
}\label{sec2}

The hairy Schwarzschild spacetime obtained by gravitational
decoupling~\cite{bib:bh1}, which satisfies all energy
conditions~\cite{bib:hok, bib:pois}, has the following form:
\begin{equation} \label{eq:metric}
ds^{2}=-e^{\mu} dt^{2}+
e^{-\lambda}dr^{2}+r^{2}
d\theta^{2}+r^{2} \sin
\theta^{2} d\varphi^{2} \,,
\end{equation}
where the metric coefficients are:
\begin{equation}
e^{\mu} = e^{-\lambda} = 1 - \frac{2M}{r}+\alpha \exp\left(
\frac{-r}{(\sigma-\alpha l/2)}
\right) \,.
\end{equation}

Here, $\alpha$ is the coupling constant, $l$ is a new charge (hair) of a
black hole~\cite{bib:bh1}, $\sigma$ is the parameter related to the
Misner--Sharp mass,  and~ $M$ is a mass of a black hole, which is given by:
\begin{equation}
M = \mathbf{M}+ \frac{\alpha l}{2} \,.
\end{equation}

$\mathbf{M}$ is the usual Schwarzschild mass. The~kinematic properties
of the solution \eqref{eq:metric} has been intensively
studied in Ref.~\cite{bib:geod}. Moreover, the~authors showed that parameters
$\alpha$ and $l$ can mimic the Kerr spacetime and gave the numerical
values for the supermassive black holes at Ark 564 and NGC 1365. The~
influence of a primary hair on the thermodynamics of a black hole
\eqref{eq:metric} has been investigated in Ref.~\cite{bib:r6}.
As we have pointed out in the introduction, gravitational
decoupling allows us to consider the Einstein equations for each source
separately; however, the~Schwarzschild solution is the vacuum solution. That
is, to obtain~\eqref{eq:metric} one should put $\tilde{T}_{ik}=0$ and
consider how an additional source $\Theta_{ik}$ changes the vacuum
Schwarzschild metric. So, the~metric \eqref{eq:metric} is the solution
of the following \mbox{Einstein equation:}
\begin{equation}
G_{ik}=-\alpha\Theta_{ik} \,
\end{equation}
where the energy-momentum tensor $\Theta_{ik}$ represents anisotropic
fluid. This energy-momentum tensor satisfies the strong and dominant
energy condition for $r\geq 2M$~\cite{bib:bh1}. It has the following
form:
\begin{equation} \label{eq:emsh}
\begin{split}
p_t=\Theta^2_2=\frac{\left(\alpha  l+r-2 \sigma \right) \alpha \exp
(\frac{2 r}{\alpha  l-2 \sigma})}{4 r \pi  \left(\alpha  l-2 \sigma
\right)^{2}}   \,, \\
P_r=-\rho=\Theta^1_1=-\frac{\alpha \exp
(\frac{2 r}{\alpha  l-2 \sigma}) \left(\alpha  l+2 r-2 \sigma \right)}{8 \left(\alpha  l-2
\sigma \right) r^{2} \pi}.
\end{split}
\end{equation}

Here, $\rho$ is the energy density of an additional matter source and
$P_r$ and $P_t$ are the radial and tangential pressure, respectively.

To obtain the line element \eqref{eq:metric} in Eddington--Finkelstein
coordinates one should perform the following coordinate transformation~\cite{bib:r6}
\begin{equation}
{dt} = {dv}+\frac{r \mathit{dr}}{\left(-\alpha
{\mathrm e}^{-\frac{2 r}{-\alpha l+2 M}} r+2 M-r\right)}
\end{equation}

Then, one obtains the hairy Schwarzschild spacetime in
Eddington--Finkelstein coordinates:
\begin{equation} \label{eq:efmetric}
ds^{2}=-\left(1-\frac{2M
}{r}+\frac{\alpha}{{ \exp}\left(\frac{r}{-\alpha
l+2 \sigma}\right)^{2}}\right)dv^{2}+
2dvdr+r^{2} d\theta^{2}+r^{2} \sin
\theta^{2} d\varphi^{2}
\end{equation}

We know that the energy-momentum tensor of the generalized Vaidya
spacetime represents the mixture of two matter fields---type I (the null
dust) and type II (the null string)~\cite{bib:vunk}. We can obtain type I if we assume that the mass function $M$ depends upon the time $v$, and~we can acquire type II if the mass function depends on the radial coordinate $r$; furthermore, we can obtain their combination if the mass function is the function of both $v$ and
$r$
. We begin our consideration by assuming that $M$ is the function of
the $r$ coordinate only ($M=M(r)$). With~this assumption, the~Einstein
tensor components $G_{ik}$ for the metric~\eqref{eq:efmetric} are given
by:
\begin{equation} \label{eq:g00}
G_0^0=\frac{\alpha \left(\alpha l+2 r-2 \sigma \right) \exp\left(
{\frac{2 r}{\alpha l-2 \sigma}}\right)-2 {M'(r)}
\left(\alpha l-2 \sigma \right)}{r^{2} \left(\alpha l-2 \sigma \right)}
\end{equation}
\begin{equation} \label{eq:g11}
G_1^1=\frac{\alpha \left(\alpha l+2 r-2 \sigma \right) \exp\left(
{\frac{2 r}{\alpha l-2 \sigma}}\right)-2 {M'(r)}
\left(\alpha l-2 \sigma \right)}{r^{2} \left(\alpha l-2 \sigma \right)}
\end{equation}
\begin{equation} \label{eq:g22}
G_2^2=\frac{2\alpha \left(\alpha l+2 r-2 \sigma \right) \exp\left(
{\frac{2 r}{\alpha l-2 \sigma}}\right)-
{M''(r)}
\left(\alpha l-2 \sigma \right)^2}{r \left(\alpha l-2 \sigma
\right)^2}
\end{equation}
\begin{equation} \label{eq:g33}
G_3^3=\frac{2\alpha \left(\alpha l+2 r-2 \sigma \right) \exp\left(
{\frac{2 r}{\alpha l-2 \sigma}}\right)-
{M''(r)}
\left(\alpha l-2 \sigma \right)^2}{r \left(\alpha l-2 \sigma
\right)^2},
\end{equation}
with the energy-momentum tensor:
\begin{equation}
T_{ik}=\tilde{T}_{ik}+\alpha\Theta_{ik} \,
\end{equation}
where $\Theta_{ik}$ is the energy-momentum tensor \eqref{eq:emsh} and
$\tilde{T}_{ik}$  is the energy-momentum tensor of the metric \eqref{eq:efmetric} with $\alpha=0$.
We write it in the following form:
\begin{equation} \label{eq:emrtemt}
\tilde{T}_{ik}=(\hat{\rho}+\hat{p})(l_in_k+n_il_k)+\hat{p}\tilde{g}_{ik} \,.
\end{equation}
where $\tilde{g}_{ik}$ is the metric tensor \eqref{eq:efmetric} with
$\alpha=0$. $\hat{\rho}$ and $\hat{p}$ are the energy density and the
pressure of the matter $\tilde{T}_{ik}$. $l_i$ and $n_i$ are two null
vectors, which are given by:
\begin{equation}
\begin{split}
n_{i}=\frac{1}{2}\left(1-\frac{2 M(r)}{r}\right)
{\delta^{0}_{i}}-{\delta^{1}_{i}} \,, \\
l_{i}=\delta^{0}_{i} \,, \\
l_{i}l^{i} = n_{i}n^{i} = 0 \,, \\
n_{i}l^{i} = -1 \,.
\end{split}
\end{equation}

First of all, let us find the Einstein equation in the case  $\alpha = 0$ :
\begin{equation}
\begin{split}
\hat{\rho}=-\frac{2M'(r)}{r^2} \,, \\
\hat{p}=\frac{M''(r)}{r} \,.
\end{split}
\end{equation}

The Einstein tensor components, in~this case, are given by:
\begin{equation} \label{eq:emtdm}
\begin{split}
\tilde{G}_0^0=-\frac{2 {M'(r)}}{r^{2}} \,, \\
\tilde{G}_1^1=-\frac{2 {M'(r)}}{r^{2}} \,, \\
\tilde{G}_2^2=-\frac{{M''(r)}}{r} \,, \\
\tilde{G}_3^3=-\frac{{M''(r)}}{r} \,.
\end{split}
\end{equation}

Comparing \eqref{eq:g00}, \eqref{eq:g11}, \eqref{eq:g22}, \eqref{eq:g33}
and \eqref{eq:emtdm}, one can easily decouple the initial Einstein tensor
into $\tilde{G}_{ik}$,  ($\alpha=0$), and~$\hat{G}_{ik}$, which correspond
to  the metric of minimal geometric deformation. So, one has:
\begin{equation} \label{eq:emrt}
\begin{split}
\hat{G}_0^0=\rho=-P_r=\frac{\alpha \left(\alpha l+2 r-2 \sigma \right)
{\exp}\left({\frac{2 r}{\alpha l-2 \sigma}}\right)}{r^{2} \left(\alpha
l-2\sigma\right)} \,, \\
\hat{G}_1^1=\frac{\alpha \left(\alpha l+2 r-2 \sigma \right)
{\exp}\left({\frac{2 r}{\alpha l-2 \sigma}}\right)}{r^{2} \left(\alpha
l-2
\sigma\right)} \,, \\
\hat{G}_2^2=\frac{2\alpha \left(\alpha l+2 r-2 \sigma \right)
{\exp}\left({\frac{2 r}{\alpha l-2 \sigma}}\right)}{r \left(\alpha
l-2 \sigma\right)^2} \,, \\
\hat{G}_3^3=\frac{2\alpha \left(\alpha l+2 r-2 \sigma \right)
{\exp}\left({\frac{2 r}{\alpha l-2 \sigma}}\right)}{r \left(\alpha
l-2 \sigma\right)^2} \,.
\end{split}
\end{equation}

The energy-momentum tensor must satisfy the conservation equation, which
is automatically satisfied through the Einstein equation:
\begin{equation}
T^{ik}_{;k}=\tilde{T}^{ik}_{;k}+\alpha \Theta^{ik}_{;k}=0 \,.
\end{equation}

Hence, for~two sources, one has either energy exchange between two matter
fields:
\begin{equation}
\tilde{T}^{ik}_{;k}=-\alpha \Theta^{ik}_{;k} \neq 0 \,,
\end{equation}
or purely the gravitation interaction of two sources:
\begin{equation}
\tilde{T}^{ik}_{;k}=\alpha \Theta^{ik}_{;k}=0 \,.
\end{equation}

The last condition means that $\Theta_{ik}$ corresponds to the dark matter
due to only gravitational interaction.  In~our case, from~the condition
$\tilde{T}^{ik}_{;k}=0$, it follows that $\Theta^{ik}_{;k}=0$, i.e.,~there
is no energy exchange between two sources. Let us introduce the
generalized density $\tilde{\rho}$ and pressure $\tilde{P}$ for the
metric \eqref{eq:efmetric}:
\begin{equation}
\tilde{\rho}=\frac{-\alpha \left(\alpha l+r-2 \sigma \right)\exp\left(\frac{2
r}{\alpha l-2 \sigma}\right)+ 2M''(r) \left(\alpha l-2 \sigma
\right)^{2}}{8 r^{2} \pi \left(\alpha l-2 \sigma \right)^{2}}
\end{equation}
\begin{equation}
\tilde{P}=\frac{2 \alpha \left(\alpha l+r-2 \sigma \right) \exp\left(\frac{2
r}{\alpha l-2 \sigma}\right)-M''(r) \left(\alpha l-2
\sigma
\right)^{2}}{8 r \pi \left(\alpha l-2 \sigma \right)^{2}}
\end{equation}

It is worth noticing that this decoupling was introduced in Ref.~\cite{bib:gd1}.
We have transformed it to Eddington--Finkelstein
coordinates~\eqref{eq:efmetric} because it is a effective tool to obtain Vaidya
and generalized Vaidya solutions by gravitational decoupling. We also
notice that a new gravitational source $\Theta_{ik}$ changes the
location of the apparent horizon. To~prove it, let us consider the
expansion $\Theta_l$ of outgoing null geodesic congruence:
\begin{equation}
e^{\gamma}
\Theta_l=\frac{2}{r}\left(1-\frac{2M(r)
}{r}+\frac{\alpha}{{ \exp}\left(\frac{r}{-\alpha
l+2 \sigma}\right)^{2}}\right) \,.
\end{equation}

So, to~obtain the apparent horizon, one should solve the following equation:
\begin{equation} \label{eq:ehmr}
1-\frac{2M(r)
}{r}+\frac{\alpha}{{ \exp}\left(\frac{r}{-\alpha
l+2 \sigma}\right)^{2}} =0 \,.
\end{equation}

One should note that this solution is static. It means that the apparent
horizon coincides with the event horizon. In~a dynamical case, it is not
true, and~the location of the event horizon is the big question. The~
only thing that we know is that in a dynamical case, the~radius of the
apparent horizon $r_{ah}$ is bigger than the event horizon location
$r_{eh}$ $(r_{ah} \geq r_{eh})$. The~horizon of this metric is a canonical
one~\cite{bib:wisser} if the following condition is held:
\begin{equation}
\frac{dB}{dr}|_{r=r_h}<1 \,.
\end{equation}
where $r_h$ is the solution of \eqref{eq:ehmr} and
\begin{equation}
B(r)\equiv 2M(r)-r \frac{\alpha}{{ \exp}\left(\frac{r}{-\alpha
l+2 \sigma}\right)^{2}} \,.
\end{equation}

To understand the structure of a singularity of a new solution, one
should count the Kretschmann scalar $K=R_{iklm}R^{iklm}$. Here, we do not
investigate the question of the global structure of this singularity.
The main question, which we are interested in now, is that a new
solution does not generate new singularities except for $r=0$. In~the
next section, the~singularity location will be at $r=0$ only due to the
fact that the mass function $M$ depends only on the time $v$. However,
when $M$ is the function of $r$, the~structure of the point $r=0$ is not
so clear. For~example, when one considers the generalized Vaidya
solution (without a hair), $r=0$ is not always a singular
point~\cite{bib:regular}. The~Kretschmann scalar for metric
\eqref{eq:metric} with $M=M(r)$ is given by:
\begin{equation}
\begin{array}{r}
K=\frac{1}{\left(\alpha l -2 \sigma \right)^{4} r^{6}} \bigg(-16 r \big(r^{4} M''(r)+2 r^{2} \left
(\alpha l -r -2 \sigma \right) M'(r)+\\
+\left(2 r^{2}+\left(-2 \alpha l +4 \sigma \right) r
+\left(\alpha l -2 \sigma \right)^{2}\right) M(r) \big) \alpha \left
(\alpha l -2 \sigma \right)^{2} \exp\left(\frac{2 r}{\alpha l -2
\sigma}\right)+ \\ 
+4 r^{2} \left(2 r^{2}+\left(\alpha l -2 \sigma \right)
^{2}\right)^{2} \alpha^{2} \exp\left(\frac{4 r}{\alpha l -2
\sigma}\right) + \\
+4 \left(\alpha l -2 \sigma \right)^{4} \left(r^{4} M''(r)^{2}+\left(-4 r^{3}
M'(r)+4 r^{2} M(r) \right) M''(r)+\right. \\
\left. +8 r^{2} M'(r)^{2}-16 r M M'(r)+12
M(r)^{2}\right) \bigg) \\
\end{array}
\end{equation}

\section{Vaidya Solution by Gravitational~Decoupling}\label{sec3}

The Vaidya spacetime by gravitational decoupling is obtained by the
assumption that the mass in \eqref{eq:efmetric} is the function of the
time $v$:
\begin{equation} \label{eq:efvmetric}
ds^{2}=-\left(1-\frac{2M(v)}{r}+\frac{\alpha}{{ \exp}\left(\frac{r}{-\alpha
l+2 \sigma}\right)^{2}}\right)dv^{2}+
2dvdr+r^{2} d\theta^{2}+r^{2} \sin
\theta^{2} d\varphi^{2}
\end{equation}

 The Einstein tensor components for this metric are given by:
{ \small
\begin{equation} \label{eq:evtensor}
\begin{split}
G_0^0=\frac{\alpha \left(\alpha l+2 r-2 \sigma \right) \exp\left(
{\frac{2 r}{\alpha l-2 \sigma}}\right)}{r^{2} \left(\alpha l-2 \sigma \right)} \,, \\
G_0^1= \frac{2 \dot{M}(v)}{r^{2}} \,, \\
G_1^1=\frac{\alpha \left(\alpha l+2 r-2 \sigma \right) \exp\left(
{\frac{2 r}{\alpha l-2 \sigma}}\right)}{r^{2} \left(\alpha l-2 \sigma \right)} \,, \\
G_2^2=\frac{2\alpha \left(\alpha l+2 r-2 \sigma \right) \exp\left(
{\frac{2 r}{\alpha l-2 \sigma}}\right)}{r \left(\alpha l-2 \sigma
\right)^2} \,, \\
G_3^3=\frac{2\alpha \left(\alpha l+2 r-2 \sigma \right) \exp\left(
{\frac{2 r}{\alpha l-2 \sigma}}\right)}{r \left(\alpha l-2 \sigma
\right)^2} \,.
\end{split}
\end{equation}}

Here, the~decoupling is quite simple. First of all, one can see that the
Einstein tensor $\hat{G}^i_k$ is the same as in the previous
section~\eqref{eq:emrt} and the only non-vanishing component of the Einstein
tensor $\tilde{G}^i_k$, which corresponds with case $\alpha=0$, is
\begin{equation}
\tilde{G}_0^1= \frac{2 \dot{M}(v)}{r^{2}}=-\mu \,.
\end{equation}

Here, $\mu$ is the energy density. The~energy-momentum tensor
$\tilde{T}_{ik}$ represents null dust:
\begin{equation} \label{eq:nuldust}
\begin{split}
\tilde{T}_{ik}=\mu L_iL_k \,, \\
L_i=\delta^0_i \,.
\end{split}
\end{equation}

Such as in the previous case, we have only gravitational interaction between
two matter sources:
\begin{equation}
\tilde{T}^{ik}_{;k}=\Theta^{ik}_{;k}=0 \,.
\end{equation}

Now, we calculate the expansion $\Theta_l$ in order to obtain the
apparent horizon equation:
\begin{equation}
e^{\gamma}
\Theta_l=\frac{2}{r}\left(1-\frac{2M(v)
}{r}+\frac{\alpha}{{ \exp}\left(\frac{r}{-\alpha
l+2 \sigma}\right)^{2}}\right) \,.
\end{equation}

As in the previous section, the~apparent horizon equation is:
\begin{equation}
1-\frac{2M(v)
}{r}+\frac{\alpha}{{ \exp}\left(\frac{r}{-\alpha
l+2 \sigma}\right)^{2}}=0 \,.
\end{equation}

The singularity location in the metric \eqref{eq:efvmetric} is at $r=0$,
which can be seen from the Kretschmann scalar:
\begin{equation}
\begin{split}
K=\frac{1}{\left(\alpha l -2
\sigma \right)^{4} r^{6}} \bigg(-16\left(\alpha^{2} l^{2}-2 l \left(r+2
\sigma
\right) \alpha +2 r^{2}+4 r \sigma +4 \sigma^{2}\right) \times \\
\times \, r \left(\alpha l-2 \sigma \right)^{2} \alpha M(v)
\exp\left(\frac{2 r}{\alpha l-2 \sigma}\right)+\\
+4 r^{2} \alpha^{2}
\left(\alpha^{2} l^{2}-4 \alpha l \sigma +2 r^{2}+4
\sigma^{2}\right)^{2} \exp\left(\frac{4 r}{\alpha l-2
\sigma}\right)+48 M(v)^{2} \left(\alpha l-2 \sigma
\right)^{4}\bigg)
\end{split}
\end{equation}

\section{Generalized Vaidya Spacetime by Gravitational~Decoupling}\label{sec4}

Finally, if~we consider the mass in \eqref{eq:efmetric} as the function of
both time $v$ and the space coordinate $r$, we obtain the generalized
Vaidya spacetime by gravitational decoupling:
\begin{equation} \label{efgvmetric}
ds^{2}=-\left(1-\frac{2M(v,r)
}{r}+\frac{\alpha}{{ \exp}\left(\frac{r}{-\alpha
l+2 \sigma}\right)^{2}}\right)dv^{2}+
2dvdr+r^{2} d\theta^{2}+r^{2} \sin
\theta^{2} d\varphi^{2} \,.
\end{equation}

This metric represents the Einstein equation solution of three sources:
the null dust, the~null perfect fluid, and~new field $\Theta_{ik}$. The~
Einstein tensor is given by:
\begin{equation} \label{eq:eigvt}
\begin{split}
G_0^0=\frac{\alpha \left(\alpha l+2 r-2 \sigma \right) \exp\left(
{\frac{2 r}{\alpha l-2 \sigma}}\right)-2 \mathit{M'(v,r)}
\left(\alpha l-2 \sigma \right)}{r^{2} \left(\alpha l-2 \sigma \right)} \,, \\
G_0^1= \frac{2 \dot{M}(v,r)}{r^{2}} \,, \\
G_1^1=\frac{\alpha \left(\alpha l+2 r-2 \sigma \right) \exp\left(
{\frac{2 r}{\alpha l-2 \sigma}}\right)-2 \mathit{M'(v,r)}
\left(\alpha l-2 \sigma \right)}{r^{2} \left(\alpha l-2 \sigma \right)} \,, \\
G_2^2=\frac{2\alpha \left(\alpha l+2 r-2 \sigma \right) \exp\left(
{\frac{2 r}{\alpha l-2 \sigma}}\right)-
\mathit{M''(v,r)}
\left(\alpha l-2 \sigma \right)^2}{r \left(\alpha l-2 \sigma
\right)^2} \,, \\
G_3^3=\frac{2\alpha \left(\alpha l+2 r-2 \sigma \right) \exp\left(
{\frac{2 r}{\alpha l-2 \sigma}}\right)-
\mathit{M''(v,r)}
\left(\alpha l-2 \sigma \right)^2}{r \left(\alpha l-2 \sigma
\right)^2} \,.
\end{split}
\end{equation}

We can decouple this tensor into two: one corresponds to the $\Theta_{ik}$
matter field and, exactly as in \eqref{eq:emrt}, the~other
Einstein tensor corresponds to the energy-momentum tensor
$\tilde{T}_{ik}$, which is a mixture of the two energy-momentum tensors of
type-I and type-II \mbox{matter fields}
:
\begin{equation}
\tilde{T}_{ik}=\tilde{T}^{null dust}_{ik}+\tilde{T}^{null string}_{ik} \,.
\end{equation}

Here, $\tilde{T}^{null dust}_{ik}$ is from \eqref{eq:nuldust} and
$\tilde{T}^{null string}_{ik}$ is from  \eqref{eq:emrtemt}. The~Einstein
tensor corresponding to the case $\alpha=0$ is given by:
\begin{equation}
\begin{split}
G_0^0=-\frac{2 \mathit{M'(v,r)}}{r^{2}} \,, \\
G_0^1= \frac{2 \dot{M}(v,r)}{r^{2}} \,, \\
G_1^1=-\frac{2 \mathit{M'(v,r)}}{r^{2}} \,, \\
G_2^2=-\frac{\mathit{M''(v,r)}}{r} \,, \\
G_3^3=-\frac{\mathit{M''(v,r)}}{r} \,.
\end{split}
\end{equation}

To obtain the mass function, one should impose the equation of the state
$P=\xi \rho$. Then, the~mass function is given by:
\begin{equation}
\begin{split}
M(v,r)=C(v)+D(v)r^{1-2\xi} \,, \\
\xi\neq \frac{1}{2} \,, \xi \in [-1\,, 1] \,.
\end{split}
\end{equation}
where $C(v)$ and $D(v)$ are arbitrary functions of time $v$. The~energy
conditions for the tensor $\Theta_{ik}$ were obtained in Ref.~\cite{bib:bh1}
and are the same in this case; however, weak, strong, and~dominant energy
conditions in all three cases demand:
\begin{equation}
\mu \geq 0 \,, \hat{\rho} \geq 0 \,, \hat{\rho} \geq \hat{P} \,,
\hat{P}\geq 0
\,.
\end{equation}

The interaction between $\tilde{T}_{ik}$ and $\Theta_{ik}$ is purely
gravitational, i.e.,:
\begin{equation}
\tilde{T}^{ik}_{;k}=\Theta^{ik}_{;k}=0 \,.
\end{equation}

Calculating the expansion $\Theta_l$ for null outgoing geodesic congruence
\begin{equation}
e^{\gamma}
\Theta_l=\frac{2}{r}\left(1-\frac{2M(v,r)
}{r}+\frac{\alpha}{{ \exp}\left(\frac{r}{-\alpha
l+2 \sigma}\right)^{2}}\right) \,.
\end{equation}
one can easily see that $g_{00}=0$ is again the apparent horizon equation:
\begin{equation}
1-\frac{2M(v,r)
}{r}+\frac{\alpha}{{ \exp}\left(\frac{r}{-\alpha
l+2 \sigma}\right)^{2}}=0 \,.
\end{equation}

The Kretschmann scalar shows that for all three cases the singularity
location at $r=0$ (Here, we only show that these solutions do not
generate new singularities except for the singular point at $r=0$.
However, as~we pointed out earlier, $r=0$ might be a regular point
in the case $M=M(r)$) :
\begin{equation}
\begin{array}{r}
K=\frac{1}{\left(\alpha l -2
\sigma \right)^{4} r^{6}} \bigg(-16 r \big(r^{4} M''(v,r)+2 r^{2} \left(\alpha l -r -2 \sigma \right) M'(v,r)+\\
+\left(2 r^{2}+\left(-2 \alpha l +4 \sigma \right) r
+\left(\alpha l -2 \sigma \right)^{2}\right) M(v,r) \big) \alpha \left
(\alpha l -2 \sigma \right)^{2} \exp^{\frac{2 r}{\alpha l -2
\sigma}}+ \\ 
+4 r^{2} \left(2 r^{2}+\left(\alpha l -2 \sigma \right)
^{2}\right)^{2} \alpha^{2} \exp^{\frac{4 r}{\alpha l -2
\sigma}} +\\
+4 \left(\alpha l -2 \sigma \right)^{4} \left(r^{4} M''(v,r)^{2}+\right. \\
+\left(-4 r^{3}M'(v,r)+4 r^{2} M(v,r) \right) M''(v,r)+ \\
\left. +8 r^{2} M'(v,r)^{2}-16 r M(v,r) M'(v,r)+12
M(v,r)^{2}\right) \bigg)\\
\end{array}
\end{equation}

\section{Conclusions}\label{sec5}

In this work, using the gravitational decoupling method, we obtained
new dynamical solutions---Vaidya and generalized Vaidya spacetimes.
Despite the fact that the $g_{00}$ component of Vaidya spacetimes
depends on time, we can easily decouple two (Vaidya spacetime) or
three (generalized one) gravitational sources. Moreover, we preserve the
conservation laws for the energy-momentum tensor. It means that there is no
energy exchange between these matter fields, and~they interact only by
gravitation. This fact allows us to consider $\Theta_{ik}$ as a dark
matter source. The~results of this paper will allow us to consider the
gravitational collapse problem and how the new matter field might affect
the gravitational collapse process. In~this paper, we briefly considered
the structure of the obtained spacetimes, i.e.,~we calculated only the
apparent horizon and singularity location and proved that the apparent
horizon equation is always $g_{00}=0$ and the singularity is located at
$r=0$. 
The Vaidya metric describes a dynamical spacetime instead of a static
spacetime as the Schwarzschild or Reissner--Nordstrom metrics do. In~the
real world, astronomical bodies gain mass when they absorb radiation, and~
they lose mass when they emit radiation, which means that the space time
around them is time dependent. As~we pointed out, the~Vaidya spacetime
can be used as the simplest model of gravitational collapse.
New solutions by the gravitational decoupling method allow us to
investigate the question of how an additional matter field will affect the
gravitational collapse process. When we consider the gravitational
collapse of Vaidya spacetimes, one might expect the naked
singularity to form. New solutions can tell us how $\Theta_{ik}$ will
influence the result of the gravitational collapse. Vaidya spacetimes
are currently widely used and the important question of the global
structure of new solutions is the direction of future research. We have
already explained that $\Theta_{ik}$ can be thought of as the energy-momentum
tensor of a dark matter. So, the~obtained solution can tell us how the
well-known properties of the Vaidya spacetimes change when an additional
matter field is present. These properties should also be studied in the~future.

We consider the additional matter source $\Theta_{ik}$ to be static in
this paper. However, it is interesting if one
can decouple the Einstein equations, which can be achieved if the parameter $\sigma$ connected to
the Misner--Sharp mass is also time-depended.

\textbf{ Acknowledgments}  The authors gratefully acknowledge financial
support from the RSF grant 22-22-00112. This work was performed for SAO RAS state
assignment “Conducting Fundamental Science~Research”.

\end{document}